\documentstyle[12pt]{article}
\newcommand{\Pc}{Poincar\'{e} }
\newcommand{\2}{2+1}
\title{\nopagebreak
\begin{flushright}
\tenrm UCTP104.96
\end{flushright}\vskip .7in
\large \bf  The Structure of Space-Time Emerging \\ 
            from the Two-Superbody Problem \\
            in Chern Simons Supergravity}
\author{Sunme Kim\thanks{e-mail address: skim@physunc.phy.uc.edu}
and Freydoon Mansouri\thanks{e-mail address: mansouri@uc.edu} \\ 
\it \small \it
Physics
Department, University of
Cincinnati, Cincinnati, OH 45221}
\date{}

\begin{document}
\maketitle

\begin{abstract}
We show that the exact solution of the two-superbody problem 
in $N=2$ Chern Simons supergravity in \2 dimensions leads to a 
supermultiplet of space-times characterized by the two gauge
invariant
observables of the super \Pc group. The metric of this space-time
supermultiplet can be cast into the form of a spinning cone in
which the coordinates do not commute or of a spinning cone with
an
additional finite discrete dimension. Some of the interesting
features of this
universe
and their possible physical implications are discussed in the light
of a corresponding observation by Witten. 
\end{abstract}
\pagebreak

It has been known for sometime that supergravity theories in \2
dimensions can be formulated as Chern Simons gauge theories of
the
corresponding supergroups [1-4].
The main focus of the present work is to explore the physical
properties of the emerging space-time when 
supersymmetric matter is coupled to these theories in a super \Pc
gauge
invariant manner.
Although much of what we describe is applicable to any
supergravity
theory, 
for definiteness we will consider the $N=2$ Chern Simons gauge
theory of 
the super \Pc group.

It has been pointed out recently that the two-superbody problem
in
$N=2$
Chern Simons supergravity is exactly solvable [5]. 
We will give below a physical interpretation of this solution and
show
that it
possesses a supersymmetric space-time
structure [6]. In arriving at this conclusion, 
one of the issues we will address is the question of the 
asymptotic observables associated with such a
supersymmetric space-time. In
this
connection, 
Henneaux has analyzed the metrical field theory of supergravity
(without matter) in \2  
dimensions [7]. Assuming that the space-time is
asymptotically
conical, he 
concludes that, like momenta, supercharges are not
among
the 
asymptotic observables of the supergravity theory. This is a
consequence of the fact that there are no asymptotically
covariantly constant
spinors in such a conical geometry. In our case, we
couple 
supersources (superparticles) to the Chern Simons supergravity in
a
super \Pc 
invariant manner, so that our theory is not strictly a field
theory. We find, not 
surprisingly, that the supersymmetry generators are again not
among
the
asymptotic  observables in our approach. We point out, however,
that in
analogy with the \Pc 
Chern Simons gravity case [8], where the two asymptotic
observables,
i.e., mass 
and spin, are Casimir invariants of the \Pc group, the asymptotic
observables 
of the super \Pc Chern Simons supergravity must also be
identified
with the 
Casimir invariants of the super \Pc group associated with an
equivalent one-superbody state described in [5]. We show
that the emerging
supersymmetric space-time
is
characterized
by these invariants. After establishing these results, we
compare the physical consequences of the picture which emerges
with what is expected in the
framework of an observation by Witten [9,10]

To provide the reader with some background and to make contact with
previous works, we begin with a brief
discussion of supersources which are to be coupled to the Chern
Simons action for the super \Pc group and their invariants. It
will be recalled [8] that in the case
of \Pc gravity the sources(particles) can be viewed as
irreducible representations
of the 
\Pc group in the same way as these representations are used in
particle physics in $3+1$ dimensions. Similarly, we take a
superparticle(supersource)
to
be an
irreducible representation of the super \Pc group. From this
point
of view, 
a superparticle is an irreducible supermultiplet consisting of 
several \Pc states related to each other by the action of the
supersymmetry 
generators. Clearly, this can be done for a simple or an extended
supersymmetry with or without central charges. But in the
interest of explicitness, we
will consider in detail the 
$N=2$ super \Pc group. The $N=2$  super \Pc algebra in \2
dimensions can 
be written as [6]
\begin{eqnarray}
& &[ J^{a},J^{b}] =-i\epsilon^{abc}J_{c} \hspace{.35in} ;
\hspace{.31in} [ P^{a},P^{b}]=0 \nonumber  \\
& &[ J^{a},P^{b}] =-i\epsilon^{abc}P_{c} \hspace{.31in} ;
\hspace{.31in} [ P^{a},Q_{\alpha}] =0  \nonumber \\
& &[ J^{a},Q_{\alpha}] =-(\sigma^{a})_{\alpha}^{\;\beta}Q_{\beta}
\hspace{.16in} ; \hspace{.29in} [ P^{a},Q'_{\alpha}] =0  \\
& &[ J^{a},Q'_{\alpha}]
=-(\sigma^{a})_{\alpha}^{\;\beta}Q'_{\beta}
\hspace{.16in} ; \hspace{.3in} \{ Q_{\alpha},Q_{\beta}\} =0 
\nonumber \\
& &\{ Q_{\alpha},Q'_{\beta}\} =-\sigma^{a}_{\;\alpha\beta}P_{a}
\hspace{.18in} ;
\hspace{.3in} \{ Q'_{\alpha},Q'_{\beta}\} =0 \nonumber \\
& & a=0,1,2 \hspace{.96in} ; \hspace{.34in} \alpha=1,2 \nonumber
\end{eqnarray}
The indices of the 
two
component spinor charges $Q_{\alpha}$ and $Q'_{\alpha}$ are
raised and lowered by the antisymmetric metric
$\epsilon^{\alpha\beta}$ with $\epsilon^{12}=-\epsilon_{12}=1$.
the
$SO(1,2)$ matrices $\sigma^{a}$ satisfy the Clifford algebra
\begin{equation}
\{\sigma^{a},\sigma^{b}\}=\frac{1}{2}\eta^{ab}
\end{equation}
where $\eta^{ab}$ is the Minkowski metric with signature
$(+,-,-)$.
We also have
\begin{equation}
\sigma^{a}_{\;\alpha\beta}=(\sigma^{a})_{\alpha}^{\;\gamma}
\epsilon_{\gamma\beta}\end{equation}
It is convenient to take the matrices $\sigma^{a}$ to be
\begin{eqnarray}
 \sigma^{0}=\frac{1}{2} \left( \begin{array}{cc}
                               1 & 0  \\ 
                               0 & -1 \end{array} \right) \;\;\;
;
\;\;\;
\sigma^{1}=\frac{1}{2} \left( \begin{array}{cc}
                               0 & i   \\ 
                               i & 0 \end{array} \right)  \;\;\;
;
\;\;\;
\sigma^{2}=\frac{1}{2} \left( \begin{array}{cc}
                                   0 & 1 \\ 
                                  -1 & 0 \end{array} \right)
\end{eqnarray} 
The two Casimir operators of the super \Pc group are given by
\begin{eqnarray}
C_{1}&=&P^{2}=\eta^{ab}P_{a}P_{b}  \\
C_{2}&=&\eta^{ab}P_{a}J_{b}+\epsilon^{\alpha\beta}Q'_{\alpha}
Q_{\beta}
\end{eqnarray}
The first of these is the same as the Casimir operator of the \Pc 
subgroup, 
so that its eigenvalues can be identified with the square of the
mass of 
the superparticle. Since the Pauli-Liubanski operator (or its
square) does 
not commute with supersymmetry transformations, it must be
supplemented 
with the second term on the right hand side of equation (6) to
obtain a
super \Pc 
invariant. We will designate its eigenvalues as $mc_{2}$.

Irreducible representations of the $N=2$ super \Pc
group in \2 dimensions can be constructed along the same lines as
those in 3+1 dimensions [11]. 
We note that without loss of generality we can work in a frame in
which 
the supermultiplet is at rest. Then
the non-vanishing
anti-commutators of 
the superalgebra simplify to
\begin{equation}
\{ Q_{1},Q'_{2}\} =\{ Q_{2},Q'_{1}\} =\frac{m}{2}
\end{equation}
Thus $Q_{\alpha}$ and $Q'_{\alpha}$, $\alpha =1,2$, form a
Clifford
algebra. 
We define a Clifford vacuum state , $|\Omega >$ by the
requirement
\begin{eqnarray}
Q_{\alpha}|\Omega >=0 & ; & \alpha=1,2
\end{eqnarray}
It is easy to verify that such a state exists within every
supermultiplet 
and that it is an eigenstate of $C_{1}$ and $C_{2}$:
\begin{eqnarray}  
C_{1}|\Omega > &=& m^{2} |\Omega >  \\
C_{2}|\Omega > &=& mc_{2} |\Omega >
\end{eqnarray} 
From the definition of the Clifford vacuum state in the rest
frame
of the 
superparticle, it follows that 
\begin{eqnarray}
C_{2}|\Omega > &=& P\cdot J |\Omega > \nonumber \\
               &=& ms^{0} |\Omega > \\
               &=& ms |\Omega > \nonumber
\end{eqnarray}
where we identify the eigenvalue, $s$, of the operator $s^{0}$
with
the spin 
of the state $|\Omega >$. So, the Clifford vacuum state is a \Pc
state with 
mass $m$ and spin $s$: 
\begin{equation}
|\Omega >=|m,s>
\end{equation}
Consider, next, the states
\begin{eqnarray}
|\Omega_{1}> &=& Q'_{1} |\Omega >, \\
|\Omega_{2}> &=& Q'_{2} |\Omega >
\end{eqnarray}
and
\begin{equation}
|\Omega_{12}>=Q'_{1}Q'_{2} |\Omega >
\end{equation}
It is easy to verify that 
\begin{eqnarray}
s^{0}|\Omega_{1} > &=& (s-\frac{1}{2})|\Omega_{1}> \\
s^{0}|\Omega_{2} > &=& (s+\frac{1}{2})|\Omega_{1}>  \\
s^{0}|\Omega_{12}> &=& s|\Omega_{12}>
\end{eqnarray}
These three \Pc states together with the Clifford vacuum state
form
an 
Irreducible supermultiplet of $N=2$ super \Pc group in \2
dimensions, which we
call
a 
superparticle. Each supermultiplet is distinguished by its mass
$m$
and 
the eigenvalue $c_{2}=s$, where $s$ is the spin of the Clifford
vacuum 
state. We will refer to $c_{2}$ as the superspin of the
multiplet.

One interesting feature which distinguishes the representations
of
the 
\Pc group in \2 dimensions from those in $3+1$ dimensions is that
in \2 
dimensions the little group of time-like momenta is $SO(2)$, so
that 
the spin of a \Pc state is not limited to integer and half
integer
values 
and can be any real number. The spins of the states within a
supermultiplet are fixed once the value of $c_{2}$ is specified.
For example, for $c_{2}=\frac{1}{2}$, the resulting $N=2$
supermultiplet is a
vector 
multiplet consisting of a spin zero, two spin 1/2, and one spin
one
\Pc 
states.

For later use, we give here an explicit realization of the $N=2$
super 
\Pc algebra by extending the phase space variables $p^{a}$ and
$q^{a}$ 
of the \Pc group to their supersymmetric form:  
\begin{equation}
p^{a}\longrightarrow p^{A}\equiv \left( p^{a},p^{\alpha}\right)
\;\; ; \;\;
q^{a}\longrightarrow q^{A}\equiv \left( q^{a},q^{\alpha}\right) 
\end{equation}
 In terms of these variables, the generators of the super \Pc
algebra take the form
\begin{eqnarray}
P_{a}&=& i\partial_{a}  \;\;\; ; \;\;\;
Q_{\alpha}=-\partial_{\alpha} \;\;\; ; \;\;\;
Q'_{\alpha}=i(\sigma^{a})_{\alpha\beta}q^{\beta}\partial_{a}
\nonumber\\
J_{a}&=& \epsilon_{abc}q^{b}p^{c}+(\sigma_{a})_{\alpha}^{\;
\beta}q^{\alpha}\partial_{\beta}+s_{a}
\end{eqnarray}

We now turn to the action of the two-superbody problem [5].
For the super \Pc algebra, the connection can be written as
\begin{equation}
A_{\mu}=e_{\mu}^{\;a}P_{a}+\omega_{\mu}^{\;a}J_{a}+\chi_{\mu}^{\;
\alpha}Q_{\alpha}+\xi_{\mu}^{\;\alpha}Q'_{\alpha}
\end{equation}
Then, the Chern Simons action for the super
\Pc
group can be written as
\begin{eqnarray}
I_{cs} &=& \frac{1}{2}\int_{M}\{ \eta_{bc}[e^{b}\wedge
(2d\omega^{c}+\epsilon^{c}_{\;da}\omega^{d}\wedge\omega^{a})]
\nonumber \\
& &
-\epsilon_{\alpha\beta}[\chi^{\alpha}\wedge(d-i\sigma_{a}\omega^{
a})\psi^{\beta}+\psi^{\alpha}\wedge(d-i\sigma_{a}\omega^{a})\chi^
{\beta}]\}
\end{eqnarray}
This action is invariant under the local infinitesimal gauge
transformations
\begin{equation}
\delta A_{\mu}=\partial_{\mu}u+i\left[ A_{\mu},u \right]
\end{equation}
where
\begin{equation}
u=\rho^{a}P_{a}+\tau^{a}J_{a}+\nu^{\alpha}Q_{\alpha}+\nu'^{\alpha
}Q'_{\alpha}
\end{equation}
As in the case of \Pc gravity [8], the manifold M is
specified by
its topology 
and is not to be identified with space-time which will emerge
(see below) as an output of this gauge theory.

To couple (super)sources to this Chern Simons theory, we proceed
in
a manner similar to the way sources were coupled to the \Pc Chern
Simons theory. From the discussion of the supermultiplets given
above, we
conclude
that the
logical candidates for our supersources are the irreducible
representations 
of the $N=2$ super \Pc group. Then each supersource can be
coupled
to 
the $N=2$ Chern Simons supergravity by an action of the form
[5]
\begin{eqnarray}
I_{s}=\int_{C}d\tau\{
p_{a}\partial_{\tau}q^{a}-\epsilon_{\alpha\beta}p^{\alpha}
\partial_{\tau}q^{\beta}-t^{\mu}(e_{\mu}^{a}p_{a}+\omega_{\mu}^{a
}j_{a}
-i\epsilon_{\alpha\beta}\chi_{\mu}^{\alpha}p^{\beta} \nonumber \\
+(\sigma \cdot
p)_{\alpha\beta}\xi_{\mu}^{\alpha}q^{\beta})+\lambda_{1}(p^{2}-m^
{2})
+\lambda_{2}(c_{2}-s)\}
\end{eqnarray}
where $\tau$ is an invariant parameter along the trajectory $C$.
Also, $mc_{2}$ is an eigenvalue of the second Casimir operator of
the super \Pc group, and $s$ is the spin of the Clifford vacuum
state of the supermultiplet. The
choice of the constraint multiplying $\lambda_{2}$ is crucial in
relating the eigenvalue of the second
Casimir invariant,
$c_{2}$, of the superalgebra to the spin content of a
supermutiplet. For more than one source, one can add an
action
of this type for each source. In the presence of supersources 
the topology of the manifold is modified. 
But the field strengths still vanish outside supersources,
and
the theory is locally trivial.

It was shown in reference [5] that the exact gauge
invariant observables of the two-superbody system can be obtained
in terms of Wilson loops. They may be viewed as the Casimir
invariants of an equivalent one-superbody state, as was done
in Chern Simons gravity [8]. We will refer to these
invariants as $H$ and $C_{2}$. As we have seen above, their
eigenvalues determine mass(energy) and spin(angular momentum)
content
the supermultiplet. They
constitute
the asymptotic observables of the two-superbody system. Here we
simply quote
the expressions from which $H$ and $C_{2}$ can be computed
exactly [5,6]. Since the invariant $H$ is the same as that for the
\Pc
algebra, it is given by [8]
\begin{equation}
\cos\frac{H}{2}=\cos\frac{m_{1}}{2}\cos\frac{m_{2}}{2}-
\frac{p_{1}\cdot
p_{2}}{m_{1}m_{2}}\sin\frac{m_{1}}{2}\sin\frac{m_{2}}{2}
\end{equation}

To obtain $C_{2}$, define the quantity $|Z_{\pm}|$ according to
\begin{equation}
|Z_{\pm}|=[H^{2}\pm
2C_{2}]^{\frac{1}{2}}
\end{equation}
Also define
\begin{equation}
W_{R}(C_{0})=
(2\cos\frac{|Z_{+}|}{2}-1)(2\cos\frac{|Z_{-}|}{2}-1)
\end{equation}

Then in terms of the invariants of the two-superbody system we
have
\begin{equation}
W_{R}(C_{0})=W_{x}W_{y}
\end{equation}
where
\begin{eqnarray}
W_{x} &=& \{1
 +2(\cos\frac{|x_{1}|}{2}-1)+2(\cos\frac{|x_{2}|}{2}-1) \nonumber
\\
& & -2(|x_{1}||x_{2}|)^{-1}(x^{a}_{1}\cdot
x^{a}_{2}-\epsilon_{\alpha\beta}q^{\alpha}_{1}q^{\beta}_{2})
\sin\frac{|x_{1}|}{2}\sin\frac{|x_{2}|}{2} \nonumber \\
& &
+(|x_{1}||x_{2}|)^{-2}(\cos\frac{|x_{1}|}{2}-1)
(\cos\frac{|x_{2}|}{2}-1) \\
& & \times
[\;|x_{1}|^{2}|x_{2}|^{2}+4i\epsilon_{abc}x^{a}_{1}x^{b}_{2}
(\sigma)_{\alpha\beta}
q^{\alpha}_{1}q^{\beta}_{2}+(x^{a}_{1})^{2}(x^{a}_{2})^{2}
\nonumber \\
& & \;\;\;\;\;\;
-2x^{a}_{1}x^{a}_{2}\epsilon_{\alpha\beta}q^{\alpha}_{1}q^{\beta}
_{2}
+\epsilon_{\alpha\beta}\epsilon_{\gamma\delta}q^{\alpha}_{1}q^
{\gamma}_{2}q^{\beta}_{1}q^{\delta}_{2}\;] \} \nonumber
\end{eqnarray}
In this expression,with $k=1,2$, and $x^{A}=(x^{a},q^{\alpha})$,
we
have
\begin{equation}
x^{a}_{k}=p^{a}_{k}+j^{a}_{k} \; \;\; ; \;\; \;
q^{\alpha}_{k}=p^{\alpha}_{k}+
j^{\alpha}_{k}
\end{equation}
The factor $W_{y}$ has exactly the same structure as $W_{x}$ with
$x^{A}
\longrightarrow y^{A}=(y^{a},q'^{\alpha})$,where
\begin{equation}
y^{a}_{k}=p^{a}_{k}-j^{a}_{k} \; \;\; ; \;\; \;
q'^{\alpha}_{k}=p^{\alpha}_{k}-
j^{\alpha}_{k}
\end{equation}

Having obtained the gauge invariant observables of the two-
superbody system, we now turn to the structure of the
corresponding space-time.
In studying the properties of this space-time, we are 
guided by the space-time structure which emerged 
from
the 
dynamics of the two-body system in \Pc Chern Simons gravity [8].
There it was shown that, by a suitable
choice of gauge, the spacial components of the relative phase
space variable $q$ of the two-body system specify a cone with the
deficit angle $H$ given by equation (26). 
In terms of the gauge fixed
variables, aside from the specific significance of $H$ and $s$ in
the present context, the expression for the line element has the
same form as is known for any spinning cone [12] :  
\begin{equation}
ds^{2}=dq'^{2}_{0}-dr^{2}-r^{2}d\phi'^{2}
\end{equation}
Or in terms of more familiar coordinates
\begin{equation}
ds^{2}=(dq^{0}-\frac{sd\phi}{2\pi}) ^{2}
-dr^{2}-\alpha^{2}r^{2}d\phi^{2}
\end{equation}
The coordinates in these equivalent expressions are related by
\begin{equation}
q'^{0}=q'^{0}(q^{0},\phi')=q^{0}-\frac{s \phi'}{2\pi \alpha}
\end{equation}
and
\begin{equation}
\vec{q'}=\left[ \exp i\tau^{0}J_{0} \right]\vec{q}
\end{equation}
where
\begin{equation}
\tau^{0}=(1-\frac{H}{2\pi})\equiv \alpha \phi=\phi'
\end{equation}

In the supersymmetric case, the
situation 
is somewhat more complicated. To see why, we note that
in both cases we can associate our gauge invariant observables to
a 
reduced 
one-(super)body state. In the pure gravity case, such a state is
a
single \Pc state, but in the supersymmetric case it is a
supermultiplet
consisting of 
several (four for $N=2$) \Pc states. As stated above, in the case
of \Pc Chern
Simons theory, 
the structure of the emerging space-time and
its
asymptotic 
observables are completely determined by the numerical values of
the two (gauge invariant) 
Casimir invariants of the reduced one-body \Pc state. To see how
this 
picture generalizes for the two-superbody system, we recall that
our two supersources are characterized by charges
$(p_{1}^{A},j_{1}^{A})$ and 
$(p_{2}^{A},j_{2}^{A})$ with the corresponding canonical
coordinates 
$q_{1}^{A}$ and $q_{2}^{A}$, respectively. Without loss of
generality, 
let the first supersource be at rest at the origin, i.\ e.\ ,
$\vec{q}_{1}=0$.
Then $\vec{q}_{2}\equiv \vec{q}$ can be viewed as a relative
coordinate. 
As in pure gravity, we parametrize $\vec{q}$ by its polar
components:
$\vec{q}=(r,\phi)$.
By fixing $\vec{q}_{1}=0$, we have again made a choice of gauge
which 
fixes all the $N=2$ super \Pc gauge transformations except for
the 
spatial rotations generated by $J^{0}$ and translations along
$q^{0}$. 
To fix these, consider first the same transformation as specified
by equations (36) and (37). 
Being an element of $N=2$ super \Pc group, this transformation
leaves the 
Casimir invariants $H$ and $c_{2}$ unchanged. But again the
$\vec{q'}$ is 
no longer $2\pi$ periodic and satisfies the matching conditions
for the 
coordinates on a cone characterized by the deficit angle $\beta
=H$.

Up to this point, everything looks the same as in \Pc gravity
space-time. 
However, essential differences appear when we try to gauge fix
the 
translations along $q^{0}$. It will be recalled from our
discussion of supersources
that
an $N=2$ 
supermultiplet at rest with Casimir invariants $H$ and $c_{2}$
consists
of 
four \Pc states with the following spin eigenvalues :
\begin{eqnarray}
P\cdot J|H,c_{2},s_{1}> &=& H(c_{2}-\frac{1}{2})|m,c_{2},s_{1}>
\\ 
P\cdot J|H,c_{2},s_{2}> &=& Hc_{2}|m,c_{2},s_{2}>  \\ 
P\cdot J|H,c_{2},s_{3}> &=& Hc_{2}|m,c_{2},s_{3}>  \\ 
P\cdot J|H,c_{2},s_{4}> &=& H(c_{2}+\frac{1}{2})|m,c_{2},s_{4}>
\end{eqnarray}
Let us compare this supermultiplet with the \Pc state obtained in
the 
reduction of \Pc Chern Simons gravity. In the latter
case, it was possible to further fix the
gauge in $q^{0}$ direction by the transformation (35) which
involved the spin of the \Pc state. Clearly, this is no longer
possible for a supermultiplet consisting of \Pc states
of different
spins. This
makes
it impossible for a single metric of the form (34) or (35) to
describe all
the spin states of our equivalent one-superbody multiplet.
So, to describe all the spin states
corresponding
to our gauge invariant observables $H$ and $c_{2}$, we generalize
the usual notion of a line element
to
an ``operator line element" which acts on the supermultiplet of
\Pc states. This operator has the form
\begin{equation}
ds^{2}=(dq^{0}-\frac{Sd\phi}{2\pi}) ^{2}
-dr^{2}-\alpha^{2}r^{2}d\phi^{2}
\end{equation}
where now $S=\frac{P\cdot J}{H}$ is the spin operator of the
supersymmetry
algebra. We could also regard $H$ and hence $\alpha$ as an
operator
in the expression (42). But since $H$ is a Casimir operator and has
the
same eigenvalue for all the states of a supermultiplet, we may
replace it with its eigenvalue throughout. When this operator
line
element acts on a state of a supermultiplet, we can replace $S$
with the spin eigenvalue of that state and hence specify the
corresponding space-time. It therefore follows that the
description
of all the spin
states of the equivalent one-body supermultiplet requires a
multiplet of space-times equal in number to the dimension of the
supermultiplet (four for $N=2$). The line elements for the
members
of this space-time multiplet, with $k=1,..,4$, are given by
\begin{equation}
ds_{k}^{2}=(dq^{0}-\frac{s_{k}d\phi}{2\pi}) ^{2}
-dr^{2}-\alpha^{2}r^{2}d\phi^{2}
\end{equation}

The line element operator in equation (42) is not invariant under
supersymmetry
transformations, and it transforms in the same way as the \Pc
states within a supermultiplet. In other words, for $k=1,..,4$
the
line elements in equation (43) form an irreducible representation
of
the
$N=2$
supersymmetry and are completely determined by the asymptotic
observables $H$ and $C_{2}$. Thus, the space-time description of
the two-super-
body system coupled to the super \Pc Chern Simons action requires
not just one but a supermultiplet of space-times. One way to view
this supersymmetric space-time is to regard it as an ordinary
space-time with an additional finite
discrete dimension. It is also interesting to note that we can
still write the metric
operator (42) in same form as (33) by defining  
\begin{equation}
q'^{0}=q^{0}-\frac{S\phi}{2\pi}
\end{equation}
But then it is clear that the coordinates of such a generalized
conical geometry
will no longer be c-numbers and will not commute with each other.

It will be instructive to compare our results with the
conclusions
of 
Henneaux who analyzed metrical supergravity in \2 dimensions
[7].
Assuming 
that the space-time is asymptotically conical, he showed that
there
are 
no Killing spinors associated with the supersymmetric generators
so
that 
the supercharges cannot be asymptotic observables. From our point
of view,
the
asymptotic observables of the Chern Simons supergravity coupled
to 
(super)matter are the Casimir invariants $(H,c_{2})$ which label
the 
equivalent one-superbody state. It is, therefore, not
surprising
that 
there are no observables associated with (odd) supersymmetry
generators. It is the invariant $c_{2}$ which is an asymptotic
observable and signals the presence of supersymmetry.
Moreover, as we have seen, to realize the local supersymmetry,
one
needs not just one 
conical space-time but a supermultiplet of them spanning an
irreducible 
representation of $(N=2)$ supersymmetry. The supersymmetry
generators 
act as ladder operators relating the spinning cones within the
space-time supermultiplet. 

Let us consider some physical consequences of the
supersymmetric space-time which emerges from the above analysis
and compare them with Witten's observations [9].
Although both pictures arise in \2 dimensions, it is conceivable
that one or both might have 3+1 dimensional analogues. 
Drawing on the specific features of the conical geometry
mentioned above, Witten pointed out the interesting possibility
that in \2 dimensions one can, in principle, arrange for the
vanishing of the cosmological constant without having the equality
of
masses for members of a supermultiplet. In the supersymmetric
space-time which emerges from our formalism, supersymmetry
remains exact, and particles in a supermultiplet will have the
same mass. However, the superpartners will reside in different
members of the space-time supermultiplet, thus making their
direct detection non-trivial. From this point of view, any hope
of
detecting the superpartners of the known particles will rest on
supersymmetry remaining unbroken. Otherwise, the superpartners
will end up in disconnected universes. We emphasize that such
supersymmetric worlds may turn out to be peculiar to \2
dimensions only. But still it would be of interest to see how one
can experimentally establish or rule out the existence of such a
superworld. We hope to address this question elsewhere.

\bigskip
We would like to thank F. Ardalan for discussions and helpful
suggestions.
This work was supported in part by the Department of Energy under
the contract No.\ DOE-FG02-84ER40153.   

\vspace{1in}
\noindent{\bf References}

\begin{enumerate}
\item A. Achucarro and P.K. Townsend, {\it Phys. Lett} {\bf B180}
(1986) 89
\item E. Witten, {\it Nucl. Phys.} {\bf B311} (1988) 46 and {\bf
B323} (1989) 113
\item K. Koehler, F. Mansouri, C. Vaz, L. Witten, {\it Mod. Phys.
Lett.} {\bf A5}(1990) 935
\item K. Koehler, F. Mansouri, C. Vaz, L. Witten, {\it Nucl.
Phys.}
{\bf B341} 
(1990) 167 and{\bf B348}
(1990) 373
\item Sunme Kim and F. Mansouri, {\it Phys. Lett.} {\bf B
372}(1990) 72 
\item A preliminary version of our results were reported at {\it
The
International Conference on Seventy Years of Quantum Mechanics
and Modern Trends in Theoretical Physics}, Calcutta, India 1/29-
2/2/1996, to appear in the
Proceedings, ed. P. Bandyopadhyay
\item M. Henneaux, {\it Phys. Rev.} {\bf D29} (1984) 2766
\item F. Mansouri and M.K. Falbo-Kenkel, {\it Mod. Phys. Lett.}
{\bf A8} (1993) 2503; F. Mansouri, {\it Comm. Theo. Phys.} {\bf
4} (1995) 191
\item E. Witten, {\it Int. Jour. Mod. Phys.} {\bf A10} (1995)
1247
\item K. Becker, M. Becker, A. Strominger, {\it Phys. Rev.} {\bf
D51} (1995) R6603
\item A. Salam and J. Strathdee, {\it Forts. Phys.} {26} (1978)
57; P.G.O. Freund, Introduction to Supersymmetry, Cambridge
University Press, 1986
\item S.Deser, R.Jackiw, G. 't Hooft, {\it Ann. of  Phys. }(N.Y.)
{\bf 152} (1984) 220; 
S. Giddings, J. Abbott and K. Kuchar, {\it Gen. Rel. Grav.}
{\bf 16}(1984) 751;
J.R. Gott and M. Alpert, {\it Gen. Rel. Grav.} {\bf 16}
(1984) 751

\end{enumerate}

\end{document}